\documentclass[a4paper,conference,10pt]{IEEEtran}

\usepackage{amsmath,amsfonts,amssymb,bm}
\usepackage[tight,footnotesize]{subfigure}
\usepackage{graphicx,stfloats}
\usepackage{algorithm}
\usepackage{algorithmic}
\usepackage{cite,url}
\usepackage{cases}
\usepackage{extarrows}
\usepackage{theorem}
\usepackage{paralist}
\usepackage{psfrag}
\usepackage{subfigure}
\usepackage{bm}
\usepackage{float}
\allowdisplaybreaks[4]
\theorembodyfont{\rmfamily}

\newtheorem{Theorem}{Theorem}

\newtheorem{Lemma}{Lemma}

\newcommand{\Cset}{\ensuremath{\mathbb{C}}}
\newcommand{\Eset}{\ensuremath{\mathbb{E}}}

\makeatletter

\newcommand{\Rmnum}[1]{\expandafter\@slowromancap\romannumeral #1@}
\makeatother

\IEEEoverridecommandlockouts
\begin{document}
\title{Massive MIMO with Multi-Antenna Users: \\ When are Additional User Antennas Beneficial?}
\author{\IEEEauthorblockN{Xueru~Li$^{\dagger}$, Emil~Bj{\"o}rnson$^{*}$, Shidong~Zhou$^{\dagger}$ and Jing~Wang$^{\dagger}$}
\IEEEauthorblockA{$^{\dagger}$ State Key Laboratory on Microwave and Digital Communications\\
Tsinghua National Laboratory Information Science and Technology\\
Department of Electronic Engineering, Tsinghua University, Beijing 100084, China\\
$^{*}$ Department of Electrical Engineering (ISY), Link{\"o}ping University, SE-58183 Link{\"o}ping, Sweden.\\
Email: xueruli1206@163.com, emil.bjornson@liu.se.}
\thanks{The work is supported by National Basic Research Program (2012CB316000), National S\&T Major Project (2014ZX03003003-002), National High Technology Research, Development Program of China (2014AA01A704), National Natural Science Foundation of China (61201192), Tsinghua-Qualcomm Joint Research Program, Keysight Technologies, Inc., Key grant Project of Chinese Ministry of Education, ELLIIT and FP7-MAMMOET.}\vspace{-4ex}}
\maketitle
\begin{abstract}
We analyze the performance of massive MIMO systems with $N$-antenna users. The benefit is that $N$ streams can be multiplexed per user, at the price of increasing the channel estimation overhead linearly with $N$. Uplink and downlink spectral efficiency (SE) expressions are derived for any $N$, and these are achievable using estimated channels and per-user-basis MMSE-SIC detectors. Large-system approximations of the SEs are obtained. This analysis shows that MMSE-SIC has similar asymptotic SE as linear MMSE detectors, indicating that the SE increase from having multi-antenna users can be harvested using linear detectors. We generalize the power scaling laws for massive MIMO to handle arbitrary $N$, and show that one can reduce the multiplication of the pilot power and payload power as $\frac{1}{M}$ where $M$ is the number of BS antennas, and still notably increase the SE with $M$ before reaching a non-zero asymptotic limit.
Simulations testify our analysis and show that the SE increases with $N$. We also note that the same improvement can be achieved by serving $N$ times more single-antenna users instead, thus the additional user antennas are particular beneficial for SE enhancement when there are few active users in the system.
\end{abstract}

\section{Introduction}
Massive multiple-input multiple-output (MIMO) is a wireless multi-user communication technology that has attracted huge research interest the last few years. By employing hundreds of antennas at the base station (BS) and serving tens of users in each cell simultaneously, a drastic increase in SE can be achieved and simple coherent linear processing techniques are near optimal~\cite{Marzetta10,Hien13,Hoydis2013}. Therefore, massive MIMO is one of the key technologies for the next generation of wireless networks.

Existing studies of massive MIMO focus on single-antenna user devices~\cite{Marzetta10,Hien13,Hoydis2013}.
However, contemporary user devices already feature multiple antennas in order to boost the SE of the network as well as the users~\cite{Gesbert2007}. Since many devices (e.g., laptops and vehicles) have moderate physical sizes, the deployment of five or ten antennas per device is highly realistic, particularly for systems that operate at millimeter wave frequencies~\cite{Swindlehurst2014}. 
It is necessary to conduct performance analysis for massive MIMO systems with multi-antenna users, to understand how the additional antennas should be used. Extensive capacity analysis has been conducted for small-scale MIMO systems with multi-antenna users, but mainly with perfect channel state information (CSI)~\cite{Jindal2005,Taesang2006}. The papers~\cite{Yoo2006,Layec2008,Musavian2007} account for imperfect CSI in point-to-point and multiple access MIMO systems, but no large-system analysis is provided to study the massive MIMO behavior. For a fixed CSI estimation overhead, \cite{Bjornson2013b} claimed that it is better to serve many single-antenna users than fewer multi-antenna users. This claim has not been validated in massive MIMO.

In this paper, we analyze the SE of a massive MIMO system with estimated CSI and any number of antennas, $N$, per user. Lower bounds on the sum capacity are derived for the uplink and downlink, which are achievable by per-user-basis MMSE-SIC (minimum mean-squared error successive interference cancellation) detectors and only uplink pilots. Large-system approximations of the lower bounds are further obtained, which are tight as $M$ grows large. Furthermore, we generalize the power scaling laws from \cite{Hien13,Hoydis2013} to handle arbitrary $N$.
The analysis shows that equipping users with multiple antennas can greatly enhance the SE, particularly in lightly loaded systems where there are too few users to exploit the full multiplexing capability of massive MIMO with $N=1$, and the benefits can harvested by linear processing.

\section{System Model}
We consider a single-cell system in time division duplex (TDD) mode where the BS has $M$ antennas and serves $K$ users within each time-frequency coherence block. Each user is equipped with $N$ antennas. We assume that each coherence block contains $S$ transmission symbols and the channels of all users remain unchanged within each block. Let ${\bf G}_{k} \in \Cset^{M \times N}$ denote the channel response from user $k$ to the BS within a coherence block. The fading can be spatially correlated, due to insufficient spacing between antennas and insufficient scattering in the channel. We use the classical Kronecker model to describe the spatial correlation~\cite{Kermoal2002}:
\begin{equation}
{\bf G}_{k}={\bf R}_{r,k}^{\frac{1}{2}}{\bf G}_{w,k}{\bf R}_{t,k}^{\frac{1}{2}},
\end{equation}
where entries of ${\bf G}_{w,k} \in \Cset^{M \times N}$ follow independent and identically distributed (i.i.d.) zero-mean circularly symmetric complex Gaussian distributions.  ${\bf R}_{t,k}\in \Cset^{N \times N}$ represents the spatial correlation at user $k$ and  ${\bf R}_{r,k}\in \Cset^{M\times M}$ describes the spatial correlation at the BS for the link to user $k$.
The large-scale fading parameter is included in ${\bf R}_{r,k}$ and can be extracted as $\frac{1}{M}{\rm{tr}}({\bf R}_{r,k})$. Let ${\bf R}_{t,k} = {\bf U}_k {\bf \Lambda}_k {\bf U}_k^H$ be the eigenvalue decomposition of ${\bf R}_{t,k}$, where ${\bf U}_k \in \Cset^{N \times N}$ is a unitary matrix and ${\bf \Lambda}_k = {\rm{diag}} \{\lambda_{k,1},\cdots, \lambda_{k,N}\}$ contains the eigenvalues.

\subsection{Uplink Channel Estimation}
During the uplink pilot signalling, $B=NK$ orthogonal pilot sequences are needed to estimate all channel dimensions at the BS. Denote  the pilot matrix of user $k$ as ${\bf F}_{k}\in\Cset^{N \times B}$. Suppose each user only knows its own statistical CSI, ${\bf R}_{t,k}$, then based on~\cite{Emil2010}, the pilot matrix that minimizes the MSE of channel estimation under the pilot energy constraint ${\rm{tr}}({\bf F}_k{\bf F}_k^H)\le BP_k$ has the form of ${\bf F}_k = {\bf U}_k {\bf L}_k^{\frac{1}{2}}{\bf V}_{k}^T$, where $P_k$ is the maximum transmit power of user $k$, ${\bf L}_k = {\rm{diag}} \{l_{k,1},\cdots,l_{k,N}\}$ distributes this power among the $N$ channel dimensions, and ${\bf V}_k \in \Cset^{B\times N}$ satisfies ${\bf V}_{k}^H{\bf V}_{k} = B{\bf I}_N$ and ${\bf V}_{k}^H{\bf V}_{l} = {\bf 0}$ if $k\ne l$. Thus, the received signal at the BS is\vspace{-0.5ex}
\begin{eqnarray}
{\bf Y}= \sum\limits_{k=1}^K {\bf G}_k {\bf F}_k + {\bf N}
= \sum\limits_{k=1}^K {\bf H}_{k} {\bf D}_k^{\frac{1}{2}} {\bf V}_k^T + {\bf N}\in \Cset^{M \times B},
\end{eqnarray}
where we define ${\bf H}_{k} = {\bf R}_{r,k}^{\frac{1}{2}}{\bf G}_{w,k}{\bf U}_{t,k}$ and ${\bf D}_k = {\bf \Lambda}_{k}{\bf L}_{k}$ with $d_{k,i}$ being its $i$th diagonal element. ${\bf N}$ is the receiver noise that follows ${\rm{vec}}({\bf N})\sim {\cal{CN}}({\bf 0},\sigma^2{\bf I}_{BM})$, where ${\rm{vec}}(\cdot)$ is the vectorization operator. 
Assume that the BS knows the statistical information ${\bf D}_k$, then from \cite{Emil2010} the MMSE estimate of ${\hat {\bf h}}_k={\rm{vec}}({\bf H}_k)$ is \vspace{-0.5ex}
\begin{equation} \label{hhat}
{\hat {\bf h}}_k = \left({\bf D}_k^{\frac{1}{2}} \otimes {\bf R}_{r,k} \right)\left(\left({\bf D}_k\otimes {\bf R}_{r,k} \right)+ \frac{\sigma^2}{B}{\bf I}_{MN} \right)^{-1}{\bf b}_k,
\end{equation}
where ${\bf b}_k = {\rm{vec}}(\frac{1}{B}{\bf Y}_k {\bf V}_k^*)= {\rm{vec}}({\bf H}_k {\bf D}_k^{\frac{1}{2}} +\frac{1}{\sqrt{B}}{\bf N}{\bf V}_k^*)$ and $\otimes$ denotes the Kronecker product. Let ${\hat {\bf h}}_{k,i}$ be the $i$th column of ${\hat {\bf H}}_{k}$, then
\begin{equation}
\Eset\left\{{\hat {\bf h}}_{k,i}{\hat {\bf h}}_{k,j}^H\right\} = \left\{ \begin{array}{ll}
{\bf \Phi}_{k,i},  & {i} = {j},\\
{\bf 0},  & {i} \ne {j},
\end{array} \right.
\end{equation}
where ${\bf \Phi}_{k,i} = d_{k,i}{\bf R}_{r,k}(d_{k,i}{\bf R}_{r,k} + \frac{\sigma^2}{B}{\bf I}_M)^{-1}{\bf R}_{r,k}$.

\subsection{Uplink Achievable SE}\label{uplink}
When the receiving BS knows the perfect CSI of all users while each transmitter has only its own statistical CSI, the precoding directions of each user that maximize the sum capacity coincide with the eigenvectors of their own spatial correlation matrix~\cite{Li2010}. Let ${\bar {\bf F}}_k \in \Cset^{N \times N}$ denote the precoding matrix of user $k$ in the uplink payload data transmission phase, then ${\bar {\bf F}}_k = {\bf U}_k {\bf P}_k^{\frac{1}{2}}$, where ${\bf P}_k = {\rm{diag}}\{p_{k,1},\cdots,p_{k,N}\}$ with ${\rm{tr}}({\bf P}_k)\le P_k$ is the power allocation matrix. Although in our work, the BS is only aware of the estimated CSI, ${\bar {\bf F}}_k= {\bf U}_k {\bf P}_k^{\frac{1}{2}}$ is still a reasonable option to enhance the SE. Hence, the received signal at the BS is
\begin{eqnarray}\label{y}
{\bf y} = \sum\limits_{k=1}^K {\bf G}_k {\bar {\bf F}}_k {\bf x}_k + {\bf n} = \sum\limits_{k=1}^K {\bf H}_k {\bf \Lambda}_k^{\frac{1}{2}} {\bf P}_k^{\frac{1}{2}}{\bf x}_k + {\bf n},
\end{eqnarray}
where ${\bf x}_k \sim {\cal{CN}}({\bf 0}, {\bf I}_N)$ is the transmitted data symbol from user $k$ and ${\bf n} \sim {\cal{CN}}({\bf 0},\sigma^2 {\bf I}_M )$ is additive receiver noise.

Since the BS is only aware of the estimated CSI, the effect of the channel uncertainty on the mutual information of MIMO channels need to be addressed. For our system and signal model, we develop a lower bound on the mutual information between ${\bf x}=[{\bf x}_1,\cdots,{\bf x}_k]$ and ${\bf y}$ (with the imperfect CSI ${\hat {\bf H}} =[{\hat {\bf H}}_1,\cdots,{\hat {\bf H}}_K]$ as side-information) in the following theorem.
\begin{Theorem}\label{thrm1}
Consider the multiple access MIMO channel in~(\ref{y}), given imperfect CSI ${\hat {\bf H}} =[{\hat {\bf H}}_1,\cdots,{\hat {\bf H}}_K]$ at the BS, where ${\hat{\bf h}}_k = {\rm{vec}}({\bf H}_k)$ is given in~(\ref{hhat}). A lower bound on the mutual information  between ${\bf x}=[{\bf x}_1,\cdots,{\bf x}_k]$ and ${\bf y}$ is
\begin{eqnarray}\label{lowerbound3}
I\left({\bf y}, {\hat {\bf H}} ;{\bf x}\right)&\ge& \sum\limits_{k=1}^K \mathbb{E}\left\{ \log_2 \left|{\bf I}_N + {\bf Q}_k{\hat {\bf H}}_k^H {\bf \Sigma}_k{\hat {\bf H}}_k \right| \right\} \nonumber \\
&\triangleq& \sum\limits_{k=1}^K R_{\text{ul},k}^{\text{SIC}},
\end{eqnarray}
where ${\bf Q}_k = {\bf \Lambda}_k{\bf P}_k$, and ${\bf \Sigma}_k = (\sum\nolimits_{l\ne k} {\hat {\bf H}}_l {\bf Q}_l {\hat {\bf H}}_l^H + {\bf Z} + \sigma^2 {\bf I}_M )^{-1}$ with ${\bf Z} = \sum\nolimits_{l=1}^K\sum\nolimits_{n=1}^N \lambda_{l,n}p_{l,n}({\bf R}_{r,l} - {\bf \Phi}_{l,n})$.
The expectation is computed with respect to (w.r.t.) the channel estimates and $|\cdot|$ denotes the determinant of a matrix.
\end{Theorem}\vspace{-1ex}
{\noindent{\textit{Proof:}}} The proof is similar to that in~\cite{Yoo2006} and thus is omitted.

The capacity lower bound in Theorem~\ref{thrm1} is an SE achievable by using a per-user-basis MMSE-SIC detector while treating co-user interference as uncorrelated Gaussian noise. 
For example, with imperfect CSI at the BS, the signal ${\bf x}_k$ from user $k$ is transmitted through an effective channel ${\hat {\bf H}}_k {\bf Q}_k^{\frac{1}{2}}$, and is corrupted by ${\bf n}_{eq,k} = {\bf y}- {\hat {\bf H}}_k {\bf Q}_k^{\frac{1}{2}}{\bf x}_k$ which is uncorrelated and has ${\bf \Sigma}_k^{-1}$ as covariance matrix. Suppose ${\bf x}_k$ is chosen from a Gaussian codebook, then by applying MMSE-SIC detection to ${\bf x}_k$ and treating ${\bf n}_{{\rm{eq}},k}$ as uncorrelated Gaussian noise in the detector, we can obtain the ergodic achievable SE in Theorem~\ref{thrm1}. 

Theorem \ref{thrm1} is a generalization of the achievable SE analysis in prior works on massive MIMO~\cite{Hien13,Hoydis2013}. When $N=1$, our expression reduces to their corresponding results.

Since the SIC procedure can be computationally complex, another option is to treat the $N$ data streams as being transmitted by $N$ independent users, and use a linear MMSE detector to detect the $NK$ streams independently. Based on the same methodology as in~\cite{xueru2015UL}, the MMSE detector that maximizes the uplink SE of the $i$th stream of user $k$ is
\begin{equation}\label{mmse_rec}
{\bf f}_{k,i} = \sqrt{\lambda_{k,i}p_{k,i}}{\bf \Sigma}{\hat {\bf h}}_{k,i},
\end{equation}
where ${\bf \Sigma}=({\bf \Sigma}_k^{-1} + {\hat {\bf H}}_k {\bf Q}_k {\hat {\bf H}}_k^H)^{-1}$.  Applying  the linear detector ${\bf f}_{k,i}$ to the signal in~(\ref{y}), an uplink achievable SE of user $k$ is
\begin{equation}
R_{\text{ul},k}^{\text{MMSE}} = \sum\limits_{i=1}^N \Eset\left\{\log_2\left(1+ \eta_{k,i}^{\text{ul}} \right)\right\}
\end{equation}
where the SINR of the $i$th stream is
\begin{equation}\label{sinr_ul}
\eta_{k,i}^{\text{ul}} = \frac{\lambda_{k,i}p_{k,i}\left|{\bf f}_{k,i}^H {\hat {\bf h}}_{k,i} \right|^2}{\Eset\left\{{\bf f}_{k,i}^H\left({\bf y}{\bf y}^H -  \lambda_{k,i}p_{k,i} {\hat {\bf h}}_{k,i}{\hat {\bf h}}_{k,i}^H\right){\bf f}_{k,i} \Big| {\hat {\bf H}}\right\}}.
\end{equation}
Since interference from the user's own streams is not suppressed by ${\bf f}_{k,i}$, it is intuitive that $R_{\text{ul},k}^{\text{SIC}} \geq R_{\text{ul},k}^{\text{MMSE}}$.
\subsection{Downlink Achievable SE}\label{downlink}
To limit the estimation overhead, we assume no downlink pilot or CSI feedback from the BS to users. This is common practice in massive MIMO since only the BS needs CSI to achieve channel hardening. Hence, the users has no instantaneous CSI except to learn the average effective channel, ${\bar {\bf H}}_{k}\triangleq {\bf \Lambda}_{k}^{\frac{1}{2}} \Eset\{{\bf H}_{k}^H {\bf W}_{k}\}{\bf \Omega}_{l}^{\frac{1}{2}}$, and covariance matrix of the interference term. Let ${\bf W}_{k}\in \Cset^{M\times N}$ be the downlink precoding matrix associated with user $k$ and let ${\bf \Omega}_{k}={\rm{diag}}\{\omega_{k,i},\cdots,\omega_{k,N}\}$ allocate the total transmit power $P_k^{'}$ among the $N$ streams. Then the total transmit power from the BS is $\sum\nolimits_{k=1}^K P_k^{'}$. The received signal at user $k$ is
\begin{equation}
{\bf y}_{k} = {\bf G}_k^{H} \sum\limits_{l=1}^K {\bf W}_{l} {\bf \Omega}_{l}^{\frac{1}{2}}{\bf x}_{l} + {\bf n}_{k} \in \Cset^{N \times 1},
\end{equation}
where ${\bf x}_{l}\sim {\cal{CN}}({\bf 0},{\bf I}_M)$ is the downlink signal intended for user $l$ and ${\bf n}_{k} \sim {\cal{CN}}({\bf 0},\sigma^2{\bf I}_N)$ is the additive receiver noise. Without loss of generality, let user $k$ use ${\bf U}_k^H$ (the eigenvector matrix of its own correlation matrix) as a first step detector to adapt to the channel correlation, then the processed received signal is
\begin{equation}\label{zz}
{\bf z}_{k} = {\bf U}_k^H{\bf y}_{k} = {\bf \Lambda}_{k}^{\frac{1}{2}} {\bf H}_{k}^H \sum\limits_{l=1}^K {\bf W}_{l} {\bf \Omega}_{l}^{\frac{1}{2}}{\bf x}_{l}+{\bf U}_k^H {\bf n}_{k}.
\end{equation}
A lower bound on the mutual information $I({\bf z}_k;{\bf x}_k ) $ is developed in the following theorem. 
\begin{Theorem}\label{thrm2}
Consider the downlink signal model in~(\ref{zz}), given the average effective channel ${\bar {\bf H}}_{k}\triangleq {\bf \Lambda}_{k}^{\frac{1}{2}} \Eset\{{\bf H}_{k}^H {\bf W}_{k}\}{\bf \Omega}_{l}^{\frac{1}{2}}$ of user $k$. The mutual information between ${\bf z}_k$ and ${\bf x}_k$ is
\begin{equation}\label{lowerbound_dl}
I\left({\bf z}_k;{\bf x}_k \right) \ge \log_2 \left|{\bf I}_N + {\bar {\bf H}}_{k}^H {\bar{\bf \Xi}}_{k} {\bar {\bf H}}_{k}\right| \triangleq R_{\text{dl},k}^{\text{SIC}},
\end{equation}
where ${\bar {\bf \Xi}}_k = ({\bf \Lambda}_{k}^{\frac{1}{2}}\mathbb{E} \{ {\bf H}_{k}^H  \sum\limits_{l\ne k} ({\bf W}_{l} {\bf \Omega}_{l} {\bf W}_{l}^H) {\bf H}_{k} \}{\bf \Lambda}_{k}^{\frac{1}{2}}  + \sigma^2 {\bf I}_N)^{-1}.$
\end{Theorem} 
{\noindent{\textit{Proof:}}} See Appendix~\ref{Proof}.

The lower bound in Theorem~\ref{thrm2} can be achieved if user $k$ applies MMSE-SIC detection to ${\bf z}_k$ when regarding ${\bar {\bf H}}_{k}$ as the true channel and the uncorrelated term ${\bf z}_{k}-{\bar {\bf H}}_{k}{\bf x}_{k}$ is treated as worst-case Gaussian noise in the detector. Theorem~\ref{thrm2} generalizes the conventional SE analysis of massive MIMO from $N=1$ to arbitrary $N$.

The user can also apply a linear MMSE detector for symbol detection based on~(\ref{zz}). Denote ${\bar {\bf h}}_{k,i}$ as the $i$th column of ${\bar{\bf H}}_k$, then with knowledge of ${\bar{\bf H}}_k$ the MMSE detector for the $i$th stream of user $k$ that maximizes the corresponding downlink SE is ${\bf r}_{k,i} = {\bf \Xi}_k{\bar {\bf h}}_{k,i}$,
where ${\bf \Xi}_k = {\bar {\bf \Xi}}_k^{-1}+{\bar {\bf H}}_k{\bar {\bf H}}_k^H$. Applying ${\bf r}_{k,i}$ to~(\ref{zz}), the achievable SE of user $k$ is
\begin{equation}
R_{\text{dl},k}^{\text{MMSE}} = \sum\limits_{i=1}^N \Eset\left\{\log_2\left(1+ \eta_{k,i}^{\text{dl}}\right)\right\}
\end{equation}
where the SINR $\eta_{k,i}^{\text{dl}}$ of its $i$th stream is
\begin{equation}\label{sinr_dl}
\eta_{k,i}^{\text{dl}} = \frac{|{\bf r}_{k,i}^H{\bar {\bf h}}_{k,i}|^2}{{\bf r}_{k,i}^r \Eset\{{\bf z}_k {\bf z}_k^H \} {\bf r}_{k,i} - |{\bf r}_{k,i}^H{\bar {\bf h}}_{k,i}|^2}.
\end{equation}

Intuitively, the MMSE-SIC detector will have a higher performance than the MMSE detector in the downlink. To compare their performance in massive MIMO systems, we derive their asymptotic SEs in the large system limit in the next section.

\section{Asymptotic Analysis}
In this section, approximations of the SEs in Theorem~\ref{thrm1} and~\ref{thrm2} that are tight for large systems are derived for fixed power matrices ${\bf L}_k$, ${\bf P}_k$ and ${\bf \Omega}_k$. We consider the large system regime where $M$ and $K$ go to infinity while $N$ remains constant since the users are expected to have a relatively small number of antennas. In what follows, the notation $M \to \infty$ refers to $K$, $M \to \infty$ such that $\lim {\sup _M}{K \mathord{\left/  {\vphantom {K M}} \right. \kern-\nulldelimiterspace} M} < \infty $ and $\lim {\inf _M}{K \mathord{\left/  {\vphantom {K M}} \right. \kern-\nulldelimiterspace} M} >0$.
\vspace{-1ex}
\begin{Theorem}\label{Appro_UL_SIC}
For the uplink MMSE-SIC detector on a per-user basis, a large-system approximation of $R_{\text{ul},k}^{\text{SIC}}$ in Theorem~\ref{thrm1} is
\begin{eqnarray}\label{appr_ul}
{\bar R}_{\text{ul},k}^{\text{SIC}} &\triangleq& \sum\limits_{i=1}^N\log_2\left(1+\frac{1}{M}{\rm{tr}}\left({\bf \Phi}_{k,i}{\bf T} \right)\lambda_{k,i}p_{k,i} \right),
\end{eqnarray}
such that $R_{\text{ul},k}^{\text{SIC}} -{\bar R}_{\text{ul},k}^{\text{SIC}} \xrightarrow[M \to \infty]{} 0$, where ${\bf T}={\bf T}(\frac{\sigma^2}{M})$ is obtained by Theorem~\ref{theorem1} in Appendix A with $\rho=\sigma^2/M$, ${\bf S}={\bf Z}/M$, and ${\bf R}_b ={\lambda}_{l,i} p_{l,i}{\bf \Phi}_{l,i}$ with $b=(l-1)N+i$.
\end{Theorem}
\noindent\emph{Proof:} The main idea is to derive the large-system approximation of ${\hat {\bf h}}_{k,i}^H {\bf \Sigma}_k {\hat {\bf h}}_{k,j}$ which is the $(i,j)$th element of ${\hat {\bf H}}_k^H {\bf \Sigma}_k{\hat {\bf H}}_k$. Due to the mutual independence among the columns of ${\hat {\bf H}}_k$, only the diagonal elements of ${\hat {\bf H}}_k^H {\bf \Sigma}_k{\hat {\bf H}}_k$ remain as $M\to \infty$. \hfill{$\blacksquare$}

In comparison, the large-system SE approximation of the linear MMSE detector ${\bf f}_{k,i}$ can be derived by following the same procedures in~\cite{xueru2015UL}. The SE approximation is ${\bar R}_{\text{ul},k}^{\text{MMSE}}=\sum\nolimits_{i=1}^N\log_2(1+{\bar \eta}_{k,i}^{\text{ul}})$ where
\begin{equation}\label{appro_sinr_ul}
{\bar \eta}_{k,i}^{\text{ul}} = \frac{\lambda_{k,i}p_{k,i}\delta_{k,i}^2}{\sum\limits_{(l,n)\ne(l,i)} \lambda_{l,n}p_{l,n}\frac{1}{M}\mu_{k,i,l,n} + \frac{1}{M}\vartheta_{k,i}},
\end{equation}
where $\delta_{k,i}=\frac{1}{M}{\rm{tr}}({\bf \Phi}_{k,i}{\bf T})$, with ${\bf T}$ being given in Theorem~\ref{theorem1}.  $\mu_{k,i,l,n} = \frac{{\rm{tr}}({\bf \Phi}_{l,n}{\bf T}^{'}_{k,i})}{M(1+ \lambda_{l,n}p_{l,n}\delta_{l,n})^2}$ and $\vartheta_{k,i} =  \frac{1}{M}{\rm{tr}}({\bf \Phi}_{k,i}{\bf T}^{''})$. Both ${\bf T}^{'}_{k,i}$ and ${\bf T}^{''}$ are obtained by Theorem~\ref{theorem2} in the appendix with $\rho=\frac{\sigma^2}{M}$, ${\bf S} = \frac{{\bf Z}}{M}$, and ${\bf R}_b = {\lambda}_{l,i} p_{l,i}{\bf \Phi}_{l,i}$ ($b=(l-1)N+i$), except that ${\bf \Theta} = {\bf \Phi}_{k,i}$ for ${\bf T}^{'}_{k,i}$ and ${\bf \Theta} = {\bf Z}+\sigma^2{\bf I}_M$ for ${\bf T}^{''}$.

By comparing (\ref{appro_sinr_ul}) and Theorem~\ref{thrm1}, we can see that for the MMSE-SIC detector, the inter-stream interference of a user caused by imperfect CSI vanishes asymptotically, and only the inter-user interference remains. For the linear MMSE detector, however, the inter-stream interference $\frac{\mu_{k,i,k,n}}{M}$ remains in~(\ref{appro_sinr_ul}) as well. However, the impact of this part reduces to zero as $M$ grows. It shows that the SE improvements with multi-antenna users can be harvested in massive MIMO by linear detectors, thus a simple hardware implementation is possible.

Next, we derive large-system approximations of the downlink performance. The precoder used by the BS can be any linear precoder such as the matched filtering (MF), block-diagonal zero-forcing or MMSE precoding. Due to the limited space, we only consider the MF case:
\begin{equation}
{\bf W}_k =\frac{1}{\sqrt{\Eset\left\{{\rm{tr}}\left( {\hat {\bf H}}_{k} {\hat {\bf H}}_{k}^H \right)\right\}}}{\hat {\bf H}}_{k}.
\end{equation}
\begin{Theorem}\label{Appro_DL_SIC}
For the downlink MMSE-SIC detector and the linear MMSE detector, if the BS utilizes the MF precoder, the large-system approximations of the SEs in Theorem~\ref{thrm2} and~(\ref{sinr_dl}) are the same, which is
\begin{equation}\label{appr_dl}
{\bar R}_{\text{dl},k}^{\text{SIC}}\triangleq \sum\limits_{i=1}^N \log_2 \left(1 + \frac{\lambda_{k,i}\omega_{k,i}\frac{\alpha_{k,i}^2}{\theta_k}} {\frac{1}{M}\gamma_k\lambda_{k,i}+\frac{\sigma^2}{M} }\right),
\end{equation}
such that $R_{\text{dl},k}^{\text{SIC}} -{\bar R}_{\text{dl},k}^{\text{SIC}} \xrightarrow[M \to \infty]{} 0$, where $\theta_k = \sum\nolimits_{i=1}^N \alpha_{k,i}$, $\alpha_{k,i} = \frac{1}{M}{\rm{tr}}({\bf \Phi}_{k,i})$ and $\gamma_k=\frac{1}{M}{\rm{tr}}({\bf R}_{r,k}\sum\nolimits_{l \ne k}\sum\nolimits_{i = 1}^N \frac{\omega _{l,i}}{\theta_l} {{\bf{\Phi }}_{l,i}})$.
\end{Theorem}
\noindent\emph{Proof:} The proof is similar to Theorem~\ref{Appro_UL_SIC} and thus omitted.

Theorem~\ref{Appro_DL_SIC} shows that the SIC processing at users does not bring any advantage over the linear MMSE detector in the downlink. The reason is that ${\bar {\bf H}}_k$ is a diagonal matrix, which means that no inter-stream interference is introduced in this assumed true channel. Therefore, the SIC processing is neither necessary nor beneficial when there are no uplink pilots.
This result has positive influence on the design of user devices since it indicates low hardware requirements and simplifies the SE optimization.

\section{Power Scaling Laws}
It is shown in~\cite{Hien13,Hoydis2013} that for $N=1$, the transmit power can be reduced with retained performance as the number of BS antennas grows. Next, we generalize the fundamental result to handle any fixed $N$.

Assume the pilot power is reduced as ${\bf L}_k = \frac{1}{M^{\alpha}}{\bf L}_{k}^{(0)}$ and the payload powers are ${\bf P}_k = \frac{1}{M^{1-\alpha}}{\bf P}_{k}^{(0)}$ and ${\bf \Omega}_k = \frac{1}{M^{1-\alpha}}{\bf \Omega}_{k}^{(0)}$, where $0\le \alpha \le 1$ and the $(\bullet)^{(0)}$ matrices are fixed. We consider ${\bf R}_{r,k}=\beta_k{\bf I}_M$ where $\beta_k$ is the large-system fading of user $k$, so that the correlation matrix at the BS remains unchanged as $M$ grows. A different large-system limit is considered in this section: $M$ goes to infinity while $K$ and $N$ are fixed.
\begin{Lemma} \label{lemma_ul}
For the uplink MMSE-SIC receiver on a per-user basis, if the pilot power is reduced as ${\bf L}_k = \frac{1}{M^{\alpha}}{\bf L}_{k}^{(0)}$ and the payload powers is ${\bf P}_k = \frac{1}{M^{1-\alpha}}{\bf P}_{k}^{(0)}$, then $R_{\text{ul},k}^{\text{SIC}}-{\bar R}_{\text{ul},k}^{'} \xrightarrow[M \to \infty]{}0$ where
\begin{equation} \label{Rbar}
{\bar R}_{\text{ul},k}^{'} = \sum\limits_{i=1}^N \log_2 \left(1 + \beta_k^2\lambda_{k,i}^2 \frac{B l_{k,i}^{(0)} p_{k,i}^{(0)}}{\sigma^2 \left(z+\sigma^2\right)}\right),
\end{equation}
with $z\!=\!0$ if $0\le\alpha <1$ and $z=\sum\nolimits_{l=1}^K \beta_l{\rm{tr}}({\bf \Lambda}_l {\bf P}_{l}^{(0)})$ if $\alpha\!=\!1$.
\end{Lemma}
\noindent\emph{Proof:} The result can be obtained by deriving the SE with the power reduction, and investigating the limiting behavior of each parameter. The detailed proof is omitted.
\begin{Lemma} \label{lemma_dl}
For the downlink MMSE-SIC detector and the MMSE detector, if ${\bf L}_k = \frac{1}{M^{\alpha}}{\bf L}_{k}^{(0)}$ and ${\bf \Omega}_k = \frac{1}{M^{1-\alpha}}{\bf \Omega}_{k}^{(0)}$, then $R_{\text{dl},k}^{\text{SIC}}-{\bar R}_{\text{dl},k}^{'} \xrightarrow[M \to \infty]{}0$ where
\begin{equation} \label{Rbar2}
{\bar R}_{\text{dl},k}^{'} = \sum\limits_{i=1}^N \log_2 \left(1 + B\beta_k^2\lambda_{k,i}^2 \frac{\omega_{k,i}^{(0)} l_{k,i}^{(0)} \upsilon_{k,i}}{\sigma^2 \left(\beta_k \lambda_{k,i}\gamma+\sigma^2\right)}\right),
\end{equation}
with $\upsilon_{k,i}=\frac{\lambda_{k,i}l_{k,i}^{(0)}}{{\rm{tr}}({\bf \Lambda}_{k}{\bf L}_k^{(0)})} \in [0,1]$.
$\gamma=0$ if $0\le\alpha <1$ and $\gamma = \sum\nolimits_{l=1}^K\sum\nolimits_{i=1}^N \omega_{l,i}^{(0)}\upsilon_{l,i}$ if $\alpha=1$.
\end{Lemma}
\noindent\emph{Proof:} The result can be obtained by plugging the reduced power into Theorem~\ref{Appro_DL_SIC} and compute its limit as $M\to \infty$.

Notice that ${\bar R}_{\text{ul},k}^{'}$ and ${\bar R}_{\text{dl},k}^{'}$ are fixed non-zero values independent of $M$. Consequently, when the number of BS antennas is large enough, we can reduce the multiplication of the pilot power and the payload power as $\frac{1}{{M}}$ and achieve a non-zero asymptotic fixed SE. When $\alpha=0.5$ and $N=1$, our results reduce to the $1/\sqrt{M}$ scaling law for the pilot/payload powers proposed by~\cite{Hien13}.
\psfrag{10}[][]{\Large {10}}
\psfrag{50}[][]{\Large {50}}
\psfrag{100}[][]{\Large {100}}
\psfrag{200}[][]{\Large {200}}
\psfrag{300}[][]{\Large {300}}
\psfrag{400}[][]{\Large {400}}

\psfrag{0}[][l]{\Large {0}}
\psfrag{40}[][l]{\Large {40}}
\psfrag{80}[][l]{\Large {80}}
\psfrag{120}[][l]{\Large {120}}
\psfrag{160}[][l]{\Large {160}}
\psfrag{200}[][l]{\Large {200}}

\psfrag{N=1}{\Large {$N=1$}}
\psfrag{N=3}{\Large {$N=3$}}
\psfrag{Uplink approximation: MMSE-SIC}{\Large {Uplink approximation: MMSE-SIC}}
\psfrag{Uplink Simulation: MMSE-SIC}{\Large {Uplink Simulation: MMSE-SIC}}
\psfrag{Downlink approximation: MMSE-SIC}{\Large{Downlink approximation: MMSE-SIC}}
\psfrag{Downlink simulation: MMSE-SIC}{\Large {Downlink simulation: MMSE-SIC}}
\psfrag{Uplink/Downlink simulation: linear MMSE blablabla}{\Large {Uplink/Downlink simulation: linear MMSE}}
\psfrag{Number of BS antennas}[][cb]{\Large {Number of BS antennas}}
\psfrag{SE}[][]{\Large{Achievable sum SE (bit/s/Hz)}}

\begin{figure}[t]
\centering
\scalebox{0.47}{\includegraphics*{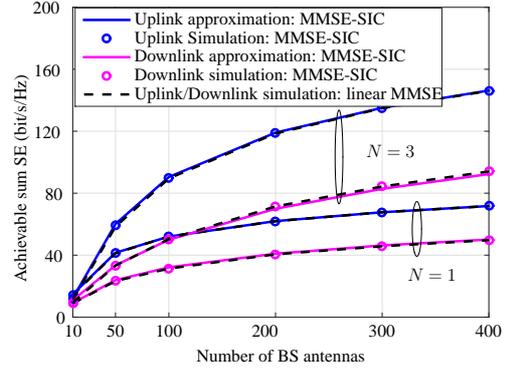}}
\caption{Uplink and downlink achievable sum SE as a function of the number of BS antennas for $K=10$.}
\label{Rate} \vspace{-1ex}
\end{figure}

\section{Simulation Results}
We consider a cell with a radius of $500$ m. The user locations are uniformly distributed at distances to the BS of at least $70$ meters. Statistical channel inversion power control is applied in the uplink, equal power allocation is used in the downlink, and the power is divided equally between the $N$ streams of each user; i.e., $\beta_ll_{l,i}\!=\!\beta_lp_{l,i}\!=\! \frac{\rho}{N\sigma^2}$ and $\omega_{l,i}\!=\! P_{d}$, where $\beta_l= \frac{1}{M} {\rm{tr}}({\bf R}_{r,l})$, with  $\rho/\sigma^2$ being set to $0\,{\text{dB}}$. $P_d$ is set to a value such that the cell-edge SNR (without shadowing) is $-3\,{\text{dB}}$.  The exponential correlation model from \cite{Loyka2001a} is used for ${\bf R}_{t,k}$ and ${\bf R}_{r,k}$. The correlation coefficients between adjacent antennas at the BS and at the users are $a_re^{j\theta_{r,k}}$ and $a_te^{j\theta_{t,k}}$, respectively, with $a_r = a_t = 0.4$, and $\theta_{r,k},\theta_{t,k}$ uniformly distributed in $[0, 2\pi)$. The coherence block length is $S=200$, which supports high user mobility.

The uplink and downlink sum SE of the MMSE-SIC and MMSE detectors are shown in Fig.~\ref{Rate}. It shows that the two detectors achieve almost the same SEs, which verifies the conclusion that a linear detector can achieve most of the SE improvements from equipping users with multiple antennas in massive MIMO. Moreover, although the pilot overhead increases, $90\%$ and $75\%$ performance gains are achieved for the uplink and the downlink, respectively, by increasing $N$ from $1$ to $3$ for $M=200$. Fig.~\ref{Rate} also verifies the tightness of the large-system approximations derived in Theorems~\ref{Appro_UL_SIC} and~\ref{Appro_DL_SIC}.

\psfrag{0}[][l]{\Large {0}}
\psfrag{20}[][l]{\Large {20}}
\psfrag{40}[][l]{\Large {40}}
\psfrag{60}[][l]{\Large {60}}
\psfrag{80}[][l]{\Large {80}}
\psfrag{10}[][l]{\Large {10}}
\psfrag{30}[][l]{\Large {30}}
\psfrag{50}[][l]{\Large {50}}
\psfrag{70}[][l]{\Large {70}}

\psfrag{$10^1$}[][]{\Large {$10^1$}}
\psfrag{$10^2$}[][]{\Large {$10^2$}}
\psfrag{$10^3$}[][]{\Large {$10^3$}}
\psfrag{$10^4$}[][]{\Large {$10^4$}}
\psfrag{$10^5$}[][]{\Large {$10^5$}}
\psfrag{$10^6$}[][]{\Large {$10^6$}}
\psfrag{$10^7$}[][]{\Large {$10^7$}}
\psfrag{Uplink SE limit}{\Large {Uplink SE limit}}
\psfrag{Uplink SE}{\Large {Uplink SE}}
\psfrag{Downlink SE limit blablabla}{\Large {Downlink SE limit}}
\psfrag{Downlink SE}{\Large {Downlink SE}}
\psfrag{alpha=0.5}{\Large {$\alpha=0.5$}}
\psfrag{alpha=1}{\Large {$\alpha=1$}}
\psfrag{M}[][cb]{\Large {Number of BS antennas}}
\begin{figure}[t]
\centering
\scalebox{0.46}{\includegraphics*{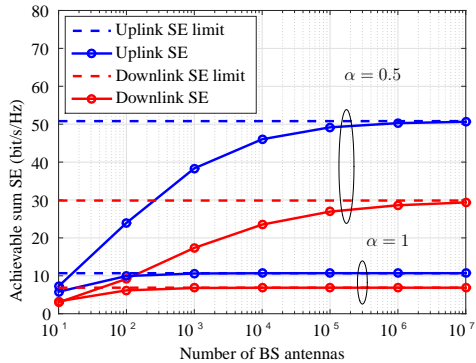}}
\caption{Power scaling law for $K=10$, $N=3$, $a_r=0$ and $a_t=0.4$.}
\label{power}
\end{figure}

Fig.~\ref{power} testifies the power scaling laws in Lemmas \ref{lemma_ul} and \ref{lemma_dl}. 
Results for $\alpha = 0.5$ and $\alpha = 1$ are shown. It is observed that, even with a $1/M$ reduction of the multiplication of pilot and payload powers, a notable increase of SE can still be obtained for an extremely wide range of $M$ before reaching the limit, especially for $M\in[50,1000]$ which is of practical interest.
\psfrag{0}[][]{\Large {0}}
\psfrag{40}[][]{\Large {40}}
\psfrag{80}[][]{\Large {80}}
\psfrag{120}[][]{\Large {120}}
\psfrag{160}[][]{\Large {160}}
\psfrag{200}[][]{\Large {200}}

\psfrag{50}[][l]{\Large {50}}
\psfrag{100}[][l]{\Large {100}}
\psfrag{300}[][l]{\Large {300}}
\psfrag{150}[][l]{\Large {150}}
\psfrag{250}[][l]{\Large {250}}
\psfrag{200}[][l]{\Large {200}}

\psfrag{N=1}{\Large {$N=1$}}
\psfrag{N=10}{\Large {$N=10$}}
\psfrag{N=3 blabla}{\Large{$N=3$}}
\psfrag{NK}[][cb]{\Large {Number of total streams $NK$}}
\psfrag{SE}[][]{\Large{Achievable sum SE (bit/s/Hz)}}
\begin{figure}[t]
\centering
\scalebox{0.46}{\includegraphics*{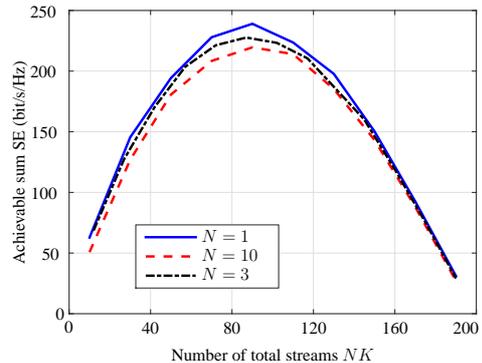}}
\caption{Achievable sum SE as a function of $NK$ for $M=200$.}
\label{ratecompare}
\end{figure}

Recall that the channel estimation overhead $NK$ equals the number of data streams that are transmitted.
For a fixed number of data streams $NK$, the system can schedule $NK$ single-antenna users and send one stream to each user, or schedule fewer multi-antenna users and send several streams to each. The downlink performance of these different scheduling approaches is compared in Fig.~\ref{ratecompare} for $N\in \{1,3,10\}$. The power per stream is $P_d$ as in Fig.~\ref{Rate}. Fig.~\ref{ratecompare} shows that for any given $NK$, scheduling $NK$ single-antenna users is always (slightly) beneficial. The optimal $NK$ is around $100$, which requires 100 active users per coherence block if $N=1$. With multi-antenna users, more realistic user numbers are sufficient to reach the sweet spot of $NK\approx 100$. Therefore, additional user antennas are beneficial to increase the spatial multiplexing in lightly and medium loaded systems.

\section{Conclusions}
We analyzed the achievable SE of single-cell massive MIMO systems with multi-antenna users. With estimated CSI from uplink pilots, lower bounds on the ergodic sum capacity were derived for both the uplink and the downlink, which are achievable by per-user MMSE-SIC detectors. Large-system SE approximations were derived and shows that the MMSE-SIC detector has an asymptotic performance  similar to the linear MMSE detector, indicating that linear detectors are sufficient to handle multi-antenna users in massive MIMO. We generalized the power scaling laws for massive MIMO from $N=1$ to arbitrary $N$. We showed that the SE increases with $N$, but for a fixed value of $NK$ the highest SE is achieved by having $NK$ single-antenna users. Hence, additional user antennas are mainly beneficial to increase the spatial multiplexing in systems with few users.

\appendices
\section{}
\begin{Theorem}($\!\!$\cite{Wagner2012}): \label{theorem1}
Let ${\bf D} \in \Cset ^{M \times M}$ and ${\bf S} \in \Cset ^{M \times M}$ be Hermitian nonnegative definite and let ${\bf H} \in \Cset ^{M \times B}$ be random with independent columns ${\bf h}_b \sim {\cal {CN}} \left(0, \frac{1}{M}{\bf R}_b \right)$. Assume that $\bf D$ and ${\bf R}_b \left( b=1,...,B\right)$, have uniformly bounded spectral norms (with respect to $M$). Then, for any $\rho > 0$,
{\small\begin{flalign} \nonumber
\frac{1}{M} {\rm{tr}}\left({\bf D} \left( {\bf {HH}}^H +{\bf S}+\rho {\bf I}_M \right)^{-1}\right)
- \frac{1}{M} {\rm{tr}}\left({\bf D}{\bf T}\left( \rho \right)\right) \xrightarrow[M \to \infty]{a.s.} 0,
\end{flalign}}
$\!\!$where ${\bf T}(\rho ) =( \frac{1}{M}\sum\nolimits_{b=1}^B \frac{{\bf R}_b}{1+\delta_b ( \rho)}+{\bf S} +\rho{\bf I}_M)^{-1}$,
and $\delta_b$ is defined as $\delta_b \left(\rho \right)=\lim_{t \to \infty}\delta_b^{\left(t \right)}\left( \rho \right)$, $ b=1,...,B$ where
{\small\begin{equation}
\delta_b^{\left(t\right)}\left( \rho \right) = \frac{1}{M} {\rm{tr}} \left({\bf R}_b \left( \frac{1}{M} \sum\limits_{j=1}^B \frac{{\bf R}_j}{1+ \delta_j^{(t-1)} (\rho )}+{\bf S} +\rho{\bf I}_N\right)^{-1}\right) \nonumber
\end{equation}}
$\!\!$for $t=1,2,\ldots,$ with initial values $\delta_b^{\left(0 \right)} = 1/\rho$ for all $b$.
\end{Theorem}
\begin{Theorem}($\!\!$\cite{Wagner2012}):\label{theorem2}
Let ${\bf {\Theta}} \in \Cset^{M \times M}$ be Hermitian nonnegative definite with uniformly bounded spectral norm (with respect to $M$). Under the same conditions  as in Theorem~\ref{theorem1},
{\small\begin{flalign}
\frac{1}{M} {\rm{tr}}\left({\bf D} {\bf A}^{-1} {\bf \Theta}{\bf A}^{-1} \right) - \frac{1}{M} {\rm{tr}}\left({\bf D}{\bf T}'\left( \rho \right) \right) \xrightarrow[M \to \infty]{a.s.} 0
\end{flalign}}
$\!\!$where ${\bf A}={\bf {HH}}^H + {\bf S} + \rho {\bf I}_M$ and ${\bf T}'( \rho ) \in \Cset^{M \times M}$ is 
{\small\begin{equation}
{\bf T}'\left( \rho \right) ={\bf T}\left( \rho \right) {\bf \Theta} {\bf T}\left( \rho \right) +{\bf T}\left( \rho \right) \frac{1}{M} \sum\limits_{b=1}^B \frac{{\bf R}_b \delta'_b\left(\rho \right)}{\left(1+ \delta_b\left(\rho \right)\right)^2}{\bf T}\left( \rho \right).
\end{equation}}
$\!\!$${\bf T}(\rho)$ and ${\delta_b}( \rho)$ are defined in Theorem~\ref{theorem1}, and ${\bm{\delta }}'( \rho)=[{\delta}'_1( \rho ),...,{\delta}'_Bt( \rho ) ]^T$ is ${\bm \delta}' (\rho)= ({\bf I}_B - {\bf J}(\rho))^{-1} {\bf v}(\rho)$ where
{\small\begin{equation}
\left[ {\bf J}\left( \rho \right)\right]_{bl} = {\frac{ \frac{1}{M} {\rm{tr}} \left({\bf R}_b {\bf T}\left( \rho \right) {\bf R}_l {\bf T}\left( \rho \right)\right)} {M \left(1+\delta_l\left( \rho \right) \right)^2 }}, 1 \le b,l \le B
\end{equation}}
{\small\begin{equation}
\left[ {\bf v}\left( \rho \right)\right]_{b} = \frac{1}{M}{\rm{tr}}\left({\bf R}_b{\bf T}\left( \rho \right) {\bf \Theta}{\bf T}\left( \rho \right)\right), 1 \le b \le B.
\end{equation}}
\end{Theorem}

\section{Proof of Theorem~\ref{thrm2}}\label{Proof}
According to the definition of mutual information, we have
\begin{equation}\label{mutual}
I({\bf z}_k;{\bf x}_k ) = h({\bf x}_k) - h({\bf x}_k|{\bf z}_k),
\end{equation}
where $h(\cdot)$ denotes the differential entropy. Then choosing the potentially suboptimal ${\bf x}_k \sim {\cal{CN}} ({\bf 0},{\bf I}_N)$ yields
\begin{equation}\label{entropy1}
h\left({\bf x}_k  \right) = \log_2\left|\pi e {\bf I}_N \right|.
\end{equation}
Meanwhile, let ${\hat {\bf x}}_k$ be the linear MMSE estimate of ${\bf x}_k$ given ${\bf z}_k$ and ${\bar {\bf H}}_{k}$, then 
${\hat{\bf x}}_k = {\bar {\bf H}}_k^H{\bf \Xi}_k{\bf z}_k$, where
{\small\begin{equation}
{\bf \Xi}_k=\left({\bf \Lambda}_{k}^{\frac{1}{2}}\Eset \left\{ {\bf H}_{k}^H  \sum\limits_{l=1}^K {\bf W}_{l} {\bf \Omega}_{l} {\bf W}_{l}^H {\bf H}_{k} \right\}{\bf \Lambda}_{k}^{\frac{1}{2}}  + \sigma^2 {\bf I}_N\right)^{-1}.
\end{equation}}
Moreover, let ${\tilde {\bf x}}_k = {\bf x}_k-{\hat {\bf x}}_k$ denote the estimation error of ${\bf x}_k$, then $h({\bf x}_k|{\bf z}_k)$ is upper bounded by the entropy of a zero-mean Gaussian vector that has the same covariance matrix as ${\tilde {\bf x}}_k$, and therefore, can be expressed as
\begin{eqnarray}\label{entropy2}
h\left({\bf x}_k|{\bf z}_k\right) &\le& \log_2\left| \pi e \Eset\left\{{\tilde {\bf x}}_k {\tilde {\bf x}}_k^H \right\}\right| \nonumber \\
&=& \log_2 \left|\pi e \left({\bf I}_N - {\bar {\bf H}}_k^H{\bf \Xi}_k{\bar {\bf H}}_k \right) \right|,
\end{eqnarray}
where the expectation $\Eset\{\cdot\}$ is with respect to the stochastic channel realizations. Plugging~(\ref{entropy1}) and~(\ref{entropy2}) into (\ref{mutual}), and applying the matrix inversion lemma, we have $I({\bf z}_k;{\bf x}_k ) \ge \log_2 |{\bf I}_N + {\bar {\bf H}}_{k}^H {\bar{\bf \Xi}}_{k} {\bar {\bf H}}_{k}|$, where ${\bar{\bf \Xi}}_{k} =( {\bf \Xi}_k^{-1} - {\bar {\bf H}}_{k}{\bar {\bf H}}_{k}^H)^{-1}$.  \hfill{$\blacksquare$}
\bibliographystyle{IEEEtran}
\linespread{1.0}\selectfont
\bibliography{ICT}
\end{document}